# The role of Dy on the magnetic properties of orthorhombic DyFeO$_3$

*Banani Biswas,[1] Veronica F. Michel,[1,5] Øystein S. Fjellvåg,[1,2] Gesara Bimashofer,[1] Max Döbeli,[3] Michal Jambor,[4] Lukas Keller,[1] Elisabeth Müller,[1] Victor Ukleev,[1] Ekaterina V. Pomjakushina,[1] Deepak Singh,[1] Uwe Stuhr,[1] C. A. F. Vaz,[1] Thomas Lippert,[1,5] and Christof W. Schneider[1,*]*

[1] Paul Scherrer Institute, 5232 Villigen PSI, Switzerland
[2] Department for Hydrogen Technology, Institute for Energy Technology, NO-2027 Kjeller, Norway
[3] Ion Beam Physics, ETH Zurich, 8093 Zurich, Switzerland
[4] Institute of Physics of Materials, Czech Academy of Sciences, Czech Republic
[5] Department of Chemistry and Applied Biosciences, Laboratory of Inorganic Chemistry, ETH Zurich, Switzerland

## Abstract

Orthoferrites are a class of magnetic materials with a magnetic ordering temperature above 600 K, predominant *G*-type antiferromagnetic ordering of the Fe-spin system and, depending on the rare-earth ion, a spin reorientation of the Fe spin taking place at lower temperatures. DyFeO$_3$ is of particular interest since the spin reorientation is classified as a Morin transition with the transition temperature depending strongly on the Dy-Fe interaction. Here, we report a detailed study of the magnetic and structural properties of microcrystalline DyFeO$_3$ powder and bulk single crystal using neutron diffraction and magnetometry between 1.5 and 450 K. We find that, while the magnetic properties of the single crystal are largely as expected, the powder shows strongly modified magnetic properties, including a modified spin reorientation and a smaller Dy-Fe interaction energy of the order of 10 μeV. Subtle structural differences between powder and single crystal show that they belong to distinct magnetic space groups. In addition, the Dy ordering at 2 K in the powder is incommensurate, with a modulation vector of 0.0173(5) *c**, corresponding to a periodicity of ~58 unit cells.

* Corresponding author: E-mail address: Christof.Schneider@psi.ch.

# I. Introduction

Materials with two or more mutually interacting ferroic orders in the same crystalline phase are known as multiferroics (MF) [1-4]. Orthoferrites ($R$FeO$_3$, $R$: rare earth element) are a well-known multiferroic class of materials with a magnetic ordering temperatures, $T_N$, above 600 K and where magnetism drives ferroelectricity. The interest in the orthoferrites resides in the various interactions within the unit cell (Fe – Fe, Fe – $R$, $R$ – $R$), which give rise to a very complex magnetic behaviour [5-7]. DyFeO$_3$ is one member of the $R$FeO$_3$ family known to have magnetic ordering-driven multiferroic properties below 4 K [8-10]. It has an orthorhombic perovskite unit cell with space group Pbnm and tilted Fe-O octahedra along the [001] direction (Fig. 1a). Below about 645 K ($T_{N,Fe}$) the Fe lattice orders along [100] in a G-type antiferromagnetic (AFM) configuration in which nearest-neighbouring Fe have antiparallel spins. The presence of a strong spin-orbit coupling leads to the presence of spin canting, resulting in a weak ferromagnetic component (wFM) along the $c$-axis, with the spin configuration represented as $\mathbf{G}_x A_y F_z$ or $\Gamma_1$ phase. The Fe-Fe interaction is dominant; however, the Fe-Dy interaction becomes important with decreasing temperature, driving a spin reorientation ($T_{SR}$) at around 50 K. The G-type ordering switches from the $a$-axis to the $b$-axis and the wFM component related to the spin canting is replaced by a C-type AFM ordering below $T_{SR}$ with the spin state represented as $A_x \mathbf{G}_y C_z$ or $\Gamma_4$ phase. Below 4 K, the Dy lattice undergoes an AFM ordering in the $ab$ plane described as $G_x A_y$. Information related to the Dy ordering is consistent in the literature, which indicates that the Dy ordering is robust [9]. However, the magnetic properties related to the Fe lattice vary, in particular $T_{SR}$, which is found to vary in the range from 37 K to 75 K [9, 11-15]. In addition, for DyFeO$_3$ nano-particles, a non-vanishing FM component below $T_{SR}$ has been reported [14, 16].

The spin reorientation in DyFeO$_3$ is related to the Dy-Fe interaction and depends on the bond distance and angle [14] and the range of values for $T_{SR}$ reported in the literature may be related to the various values found for the unit cell parameters. To understand the magnetic properties associated to the Fe-Fe and Dy-Fe interactions, we have studied the temperature-dependent magnetic properties of DyFeO$_3$ as powder and single crystal between 1.5 and 450 K using SQUID magnetometry and neutron diffraction. These measurements are complemented by structural characterization using x-ray diffraction and transmission electron microscopy diffraction, as well as room temperature x-ray absorption spectroscopy measurements at the Fe $L_{3,2}$ edge. We find that the magnetic properties of the powder differ significantly from those of the bulk crystal, which we attribute to subtle differences in the structural and electronic properties together with the sensitivity of the Dy-Fe interation to small perturbations in bond distance and angle.

## II. Experimental methods

The DyFeO$_3$ powder was prepared by a standard solid-state reaction route by mixing stoichiometric amounts of dried Fe$_2$O$_3$ and Dy$_2$O$_3$ powders (99.99 % purity), grinding, and subsequent sintering of the mixture as a pressed pellet. No single Dy-isotope enrichment has been done to correct for absorption effects in neutron powder diffraction experiments [17, 18]. After a first calcination step at 1100°C for 24 h in air, the pellet was re-grinded and sintered a second time at 1300°C for 24 h in air and cooled down at approx. 1 °C/min to ensure the correct oxygen composition. The resulting powder has an average grain size of the order of several μm, as observed using scanning electron microscopy, and the correct stoichiometry, Dy$_{1.00\pm0.02}$Fe$_{1.00\pm0.02}$O$_3$, determined using Rutherford backscattering spectrometry (RBS) [19]. From the same powder batch, the DyFeO$_3$ single crystal was grown by the floating-zone method [20] using an optical floating zone furnace (FZ-T-10000-H-IV-VP-PC, Crystal System Corp., Japan) with four 1000 W halogen lamps as a heat source. The growth rate was 5 mm/h and both rods (the feeding and seeding rod) were rotated at about 20 rpm in opposite directions to ensure the homogeneity of the melt. A mixture of argon with 10 % oxygen at 1.5 bar was applied during the growth. The crystal had the shape of a rod ~5 mm in diameter and 4 cm in length with a mosaic spread better than 1°. Single crystals of cylindrical shape and sizes were cut from the original boule, oriented with [010] along the cylindrical axis and with the [001] marked for orientation-dependent measurements.

For the structural characterization of the powder and single crystal, a Seifert HR and a Bruker D8 Advance four-circle x-ray diffractometers were used with Cu K$_{\alpha 1}$ monochromatic x-ray sources and an instrumental resolution of 0.003° calibrated with reference Si and $Al_2O_3$ single crystals. Additional powder diffraction measurements were done at the Materials Science (MS) X04SA beamline (wavelength $\lambda = 0.5653$ Å) at the Swiss Light Source (SLS, PSI Switzerland) [21]. The powder sample was placed in a 0.3 mm capillary and the experiments were carried out at room temperature. For the structural determination of the single crystal, 3D electron diffraction experiments were conducted [22] at the Electron Microscopy Facility at PSI. Thin lamellae were prepared using a Focused Ion Beam (FIB; Zeiss NVision40). A JEM-ARM200F NEOARM (JEOL Ltd., Akishima, Japan) transmission electron microscope (TEM) equipped with an *in situ* OneView camera (Gatan Inc., Pleasanton, USA), operated at 200 kV acceleration voltage, was used for the acquisition of electron diffraction patterns in a continuous rotation regime, i.e., diffraction patterns were acquired at a 20 Hz frame rate, while the specimen was continuously rotating at an angular speed of 1.73 degrees/s. The acquired electron diffraction data were processed using PETS 2.0 software and the structure was solved using JANA 2020, using a charge flipping algorithm [23, 24].

Temperature-dependent magnetic measurements were performed in a Quantum Design, MPMS® 3 SQUID magnetometer with the magnetic field aligned along the crystalline axes for the single crystal sample. Zero field (ZFC) and field cooled (FC) magnetization measurements were carried out between 2 K and 300 K with applied fields varying between 10 mT and 100 mT. Magnetization hysteresis curses (*M-H* were typically measured up to a field of 5 T. In addition, field cooled measurements upon heating or cooling (FCH, FCC) have been performed.

Neutron diffraction studies were performed at the spallation neutron source SINQ, Paul Scherrer Institute (Villigen, Switzerland). The powder neutron diffraction measurements (PND) were conducted at the cold neutron powder diffractometer DMC with incident wavelength $\lambda$ = 2.458 Å. The powder sample was placed in a double-walled Al-container and measured at 450, 315, 200, 100, 10 and 2 K in the angular range of 5–87° with 0.1° steps and data were absorption corrected for a double-walled cylinder with 10 mm outer and 8 mm inner diameter. Data analysis and Rietveld refinements were performed with software tools from the Fullprof suite [25] and JANA 2020 [23]. Symmetry analysis was carried out using the Bilbao Crystallographic Server [26, 27]. Single crystal neutron diffraction experiments were performed at the thermal triple-axis spectrometer EIGER with $k_f$ = 2.66 Å$^{-1}$ ($\lambda$ = 2.36 Å) [28] using a pyrolytic graphite filter to eliminate higher-order neutrons. The zero-field experiments were done in a He-cryostat to reach temperatures down to 1.5 K with the sample mounted in the (0*kl*) scattering plane. Experiments with magnetic fields were performed with either a vertical field pointing along [100] or a horizontal field along [010] and [001] and all field dependent measurements are ZFC measurements. Additional measurements to determine the temperature dependence of the relative change of the lattice constants of the DyFeO$_3$ single crystal were conducted at the two-axis neutron diffractometer MORPHEUS ($\lambda$ = 2.445 Å) at SINQ at 300, 200, 100, 50 and 10 K. The instrument was calibrated by measuring a Si crystal standard at room temperature.

The X-ray absorption (XAS) measurements of the powder and single crystal were performed at the SIM beamline at the Swiss Light Source [29-31]. The Fe $L_{3,2}$-edge, the O $K$-edge as well as the Dy $M_{5,4}$ XAS data were acquired at room temperature with the beam incident at 45° to the sample surface. The measurements were done in total electron (TEY) and total fluorescence yield (TFY) modes to probe the sample at different depths [32, 33]. All spectra have been normalised to the photon flux recorded by a clean gold mesh upstream of the experimental chamber. The corrected spectra were normalized using the background edge jump of the regions well below and beyond the $L_{3,2}$ (for $Fe$), $K$ (for O), and $M_{5,4}$ (Dy) absorption edges. For a qualitative analysis, a polynomial background was fitted and subtracted from the normalized spectra to emphasise the features related to oxidation states. All XAS data treatment was done using the Athena software [34].

## III. Experimental Results

### A. Crystallographic Structure

The sintered $DyFeO_3$ crystalizes in an orthorhombic structure (space group Pbnm) with lattice parameters $a = 5.304(6)$ Å, $b = 5.598(2)$ Å, and $c = 7.624(1)$ Å as determined by XRD at room temperature. These values are in good agreement with the results of the neutron powder diffraction measurements (see Table I). Figure 1(a) shows the reconstructed unit cell obtained from the XRD structural refinement. The room temperature lattice constants determined from the different techniques (XRD, ND, and TEM) for both the single crystal and powder sample are comparable, to within 95% confidence level (two standard deviations). The relative change in the lattice constants between 10 K and 300 K is similar for $b$ and $c$ (~0.2%), but significantly different for $a$, with $\Delta a =$ 0.09 % for the powder and $\Delta a$ = 0.26 % for the single crystal (see Table I ).

### B. Magnetic Structure

#### a. Magnetization measurements

The temperature dependence of the magnetic moment for the powder and the single crystal [Fig. 2 (a,b)] were measured between 300 K and 2 K. The expected Néel-temperature for the Fe in this orthorhombic structure, $T_N$~645 K, cannot be measured due to instrumental limitations. For the powder, we observe two magnetic transitions, one at ~4.7 K and one at ~54.7 K [Fig. 2(a)]. The former is identified with the ordering temperature $T_{N,Dy}$ of the Dy spins, the latter with the spin reorientation temperature $T_{SR}$ of the Fe-spins. The variation of the magnetic moment for the single crystal sample with temperature displays a pronounced orientation dependence, as shown in Fig. 2(b) (FC curves taken with 0.01 T). With increasing temperature, one finds a first transition at $T_{N,Dy}$~6 K, where a sudden increase in the magnetic moment is observed; we identify this transition with the Dy ordering temperature. Above 6 K, the magnetization along the [100] and [010] decreases, while that along [001] increases until, at 41 K, a sharp increase in the magnetization along all crystal orientations is observed, corresponding to $T_{SR}$ where the Fe-spins flip from the *b*- into the *a*-direction and the wFM fully appears. This step-like temperature dependence for the spin-flip has been reported previouly in the literature [8, 9, 11]. Above $T_{SR}$, a canting of the Fe-spins along the [001] direction gives rise to a weak FM component along the *c*-direction in DyFeO$_3$, an $F_z$ component consistent with $\Gamma_4$.

From the *M(H)* measurements, we find that for the DyFeO$_3$ powder the coercive field decreases with decreasing temperature while the remanence first increases down to $T_{SR}$ and then decreases again with lowering temperature [Fig. 3(a,b)]. A similar behavior is found for the single crystal when the field is along [001] [Fig. 3(c)], where the saturation moment reaches ~0.15 $\mu_B$/f.u. at $T_{SR}$ (where $\mu_B$ is the Bohr magnetron). The small coercive field, below 60 mT along the *c*-direction, indicates a very soft FM axis as compared to the powder sample, where the coercive field reaches 2.5 T at 150 K. For the [100] direction (see Fig. S1 in the Supplemental Materials [35]) and [010] direction [Fig. 3(d)], there is a very small or no hysteresis, demonstrating the presence of a strong uniaxial anisotropy along [001]. Counterintuitive at first sight is the decrease of the coercive field and the increase of the remanence with decreasing temperature down to $T_{SR}$. We attribute this behavior to the role of the Dy spins in assisting the magnetization reversal as they begin to order magnetically on approaching $T_{SR}$. Their impact is also observed in the increased magnetic moment as the temperature approaches $T_{SR}$ from above. Below $T_{SR}$, however, the situation changes distinctly. For the powder, the FM hysteresis persists [Fig. 3(b)] and the moment at $H = 5$ T reaches ~2.5 $\mu_B$/f.u. at 10 K and ~4.5 $\mu_B$/f.u. at 3 K. Below $T_{N,Dy}$, under an applied field, there is a switch from the $\Gamma_4$ ($A_x G_y C_z$) to the $\Gamma_1$ ($G_x A_y F_z$) phase where a FM component is expected to be present [Fig. 3(b) inset]. For the single crystal, directly below $T_{SR}$, no FM component is present for all three crystal directions but it reappears below 10 K [Fig. 3(c) inset]. The measured moment at $H = 5$ T is ~5 $\mu_B$/f.u. for the [100] direction (see Fig. S1 in the Supplemental Materials [35]), ~9-10 $\mu_B$/f.u. along [010] [Fig. 3(d)] and ~0.17 $\mu_B$/f.u. along [001]. Along the [100] and [010] directions the remanent moment vanishes, showing that DyFeO$_3$ displays dominant antiferromagnetic interations but with a small net moment along [001] [36]. The observed increase in moment below $T_{SR}$ found for both powder and single crystal is correlated with an increase in the Fe-Dy interaction with decreasing temperature and Dy ordering [6, 37].

Depending on how the magnetisation measurements are carried out (ZFC, FCC), differences in the temperature dependence for the [100] direction are observed [Fig 2(d)], with the applied field strongly modifying $T_{SR}$. In particular, a small jump at ~65 K is observed when conducting a FCC measurement with an applied magnetic field of 0.005 T, before the SR at ~41 K takes place. This is similar to the transition signature along [100] reported for DyFeO$_3$ [11] and SmFeO$_3$ [38-42]. There is also a very broad magnetic transition visible at 120 K upon heating measured at 0.02 T and ~150 K on cooling in 0.005 T and the transition temperatures seem to be influenced by the applied magnetic field [Fig 2(d)]. In addition, we find a small but clear steplike transition at ~80 K when measuring the (011) Bragg peak with neutron diffraction during a ZFC measurement [Fig. 4(a) inset]. With a small applied field or after remounting the sample, the transition was no longer noticeable. These observations show the large susceptibility of the system to small modifications with respect to the excitation conditions.

The observation of a rising slope in the moment with decreasing temperature [Figs. 2(b) and 2(d)] deviating from a simple Curie-Weiss dependence suggests that the net magnetic moments of the Fe and Dy lattice are adding up (the exact contribution of the Dy to the total moment cannot be estimated from the data). On the one side there is a paramagnetic contribution from Dy, and on the other side a contribution from the antiferromagnetically ordered Fe. The $\Gamma_4$ to $\Gamma_1$ transition requires a strong interaction between Dy and Fe since the $\Gamma_1$ state is the energetically less favourable state [37]. Therefore, the observation of an almost linear temperature dependence [Fig. 2(d)] is the result of competing Fe and Dy-moments with different temperature dependences and the jump-like transition at ~65 K is likely to be a consequence of an increase in the Dy-Fe interaction, which provokes a partial SR at a temperature higher than that at which the majority of the Fe-spins would flip.

### b. Neutron diffraction measurements

*Single crystal neutron diffraction measurements*

The SQUID measurements described above indicate that, for the single crystal, the Fe magnetic moment ($m_{Fe}$) direction is in the *ab*-plane and is expected to be along [100] above $T_{SR}$ and along [010] below, with the Dy ordering in the *ab*-plane below $T_{N,Dy}$ (Fig. 2b). To determine more precisely the magnetic structure of DyFeO$_3$ than is possible with bulk magnetometry, we turn to the results of neutron diffraction, which enable us to obtain information on both spin amplitude and direction of the different magnetic lattices [43]. Specifically, we measured the magnetic field dependence of the Fe and Dy spins in the single crystal for fields in the range from 0.01 to 1.5 T applied along the main crystalline axes. For these measurements, we selected the (011), (01-1), (031) and (03-1) magnetic Bragg peaks. Measuring (0*kl*) with odd *k*, *l* gives both Fe (0*k*0) and Dy (00*l*) magnetic contributions in one measurement and the direction of the Fe moment above $T_{SR}$ is along [100] [44].

The temperature dependence of the maximum peak intensity of the (031) and (011) Bragg peaks of the single crystal, presented in Fig. 4(a), shows that the signal above $T_{SR}$ is essentially flat indicating that all measured moments are aligned along [100]. At $T_{SR}$, there is a sharp drop of the intensity in addition to another step between 10 and 20 K before a steep increase in signal below 6 K takes place when the Dy-ordering sets in. These ZF measurements show a similar temperature dependence as the SQUID measurements [Fig. 2(b)]. The Fe-spins flip from the *a*- to the *b*-direction below $T_{SR}$ and only below 10 K most of the spins orient along [010]. For the measured temperature dependence of the (031) peak above $T_{SR}$ we find a reproducible broadening between 40 and 50 K not measured for the (011) peak whereas the drop in intensity of the (031) diffraction peak below $T_{SR}$ is sharper than that of the (011) peak. One reason why this broading of the transition above $T_{SR}$ is observed for (031) and not for (011) is the stronger weight of the Fe spins relative to Dy in the (031) reflection which indicates that some of the Fe spins start to change direction prior to the main transition. In Fig 4(b)-4(d) the temperature dependence of the intensity of the (031) reflection is presented for different magnetic field strengths and directions. These measurements were performed under a magnetic field during heating after zero field cooling (ZFC). Common for all measurements with the magnetic field applied along all three crystalline axes is that small to modest fields of up to 1.5 T are sufficient to reduce or even suppress $T_{SR}$. As pointed out, in this measurement geometry, the signal above $T_{SR}$ originates from the Fe spin along [100] since Dy is not ordered. Applying a field and hence moving the SR to a lower temperature means that the Dy-Fe interaction is reduced by the magnetic field. In particular, magnetic fields between 10 and 100 mT along [001] are already sufficient for a significant shift of $T_{SR}$ while 0.5 T results in a complete suppression of the spin reorientation. This indicates that very low energies of the order of 10 μeV or less drive the SR [33] and relatively small perturbations to the system are sufficient to modify its propertiesThe influence of the Dy moment on the Fe-spin at low temperatures and around $T_{SR}$ on the maximum intensity of the (031) peak can be seen in Figs. 4(b) and 4(c) when applying $H \parallel a$ or $b$. Up to a field of 0.5 T, $T_{SR}$ is shifted to lower temperatures by several K and the drop in intensity becomes less sharp. For the same field range, the intensity of the (031)/(03-1) Bragg peak increases at temperatures at 5 K and above with increasing field but decreases with increasing temperature up to about 30 K when reaching a minimum. This behavior indicates a competition between the Dy-induced spin flip near $T_{SR}$ and the Dy-Fe interaction at temperatures below 30 K as the magnetic moment orientation is modified by the applied magnetic field. For a magnetic field larger than 0.75 T all spins have switched from the [010] to the high temperature [100] easy axis direction of $DyFeO_3$. This is reminiscent of a metamagnetic transition, where the magnetic field flips the system towards a new energy minimum defined by the easy magnetic axis observed above $T_{SR}$, i.e., the field opposes the Dy-Fe interaction and reestablishes the Fe-Fe magnetic anisotropy term as the dominant contribution. These results confirm experimentally that the [100] direction is the energetically favourable direction for the Fe and the influence of the Dy moment causes the spin reorientation at lower temperatures.

The spin system is more susceptible to fields applied along [001], for which less energy is required to suppress the SR. In particular, we find that 10 mT is already enough to noticeably change the temperature dependence of the (031) and (011) reflections [Fig. 4(d)]. These measurements corroborate the field dependent magnetization measurements shown in Fig. 2(c). The measurements with $H \parallel c$ reveal further that this configuration is more effective in rearranging the Dy spins. To modify the Dy spin configuration with $H \parallel a$ or $H \parallel b$, an energy of at least 100 μeV is required, as shown in Figs. 4(b) and (c) [33]. The temperature dependencies of the (03-1) diffraction peaks with the applied field along the *a*- and *b*- directions are very similar [Fig.4(b) and 4(c)], but different from that along the *c*-direction [Fig.4(d)]. In contrast, for fields applied along [001], a shift of the transition temperature down to the Dy ordering temperature prevents the Fe spin reorientation altogether. The similarities of the temperature dependencies with the field along [100] or [010] are somehow surprising. Although these two directions correspond to the initial and final directions of the spins for the spin reorientation, the effect on the spin structure is very similar. This indicates that the interaction between field and Fe is indirect, probably mediated by the Dy ions.

*Powder neutron diffraction measurements*

Whereas the spin orientations for the single crystal agree with previous reports [5, 9, 38], the powder DyFeO$_3$ system shows unexpected different properties. From the powder neutron diffraction pattern [Fig. 4(e)], the extracted direction of the magnetic moments, $m_{Fe}$, above $T_{SR}$ is along [100] and flips into [001] below $T_{SR}$. When Dy orders, it orders with the spins lying in the *b-c* plane (see Table I). We also observe a peak splitting for the (001) reflection in the powder data, corresponding to incommensurate Dy ordering.

In the temperature region between $T_N$ and $T_{SR}$ (100 – 450 K), Le Bail fits of the powder data show that all reflections can be described by the nuclear unit cell, *i.e.,* the magnetic propagation vector is $k$ = (0, 0, 0). We find that the high temperature phase is described by $\Gamma_4$ (Table II), corresponding to $G_xA_yF_z$, and that the $G_x$ component dominates while the two other components do not influence the fit and therefore were set to zero for the refinements. We also note that the direction of the $F_z$ component is in agreement with the wFM observed in the single crystal. The derived model is thus a G-type AFM with moments along [100] as illustrated in Fig. 1(b) and described in Table II. Refined magnetic moments are given in the Supplemental Materials [35]. Below $T_{SR}$, the magnetic structure changes from $\Gamma_4$ to $\Gamma_2$ (Table II) and the ordering is described as $F_xC_yG_z$, evident from the change in intensity of the (011) and (101) reflections at 1.39 - 1.45 Å$^{-1}$ [Fig. 4(b)]. The implication is that the spins have flipped from [100] to [001] while keeping the G-type magnetic ordering consistent with a Morin transition. The amplitudes of $F_x$ and $C_y$ were set to zero for the refinements.

The difference curve between 2 K and 10 K [Fig. 4(f)] highlights the magnetic changes associated with the Dy ordering below 6 K [Fig. 4(f)]. We identify the strong (001) reflection arising from $A_y$ ordering with ferromagnetic layers and moments along [010] in the *ab*-plane stacked antiferromagnetically along [001], corresponding to $\Gamma_5$ for Dy (Table II). However, $\Gamma_5$ does not give any intensity for the (010) reflection at 1.15 Å$^{-1}$, and the (101) and (011) reflections are not well described either. By adding a $G_z$ component corresponding to $\Gamma_7$, we successfully describe the data (Fig. 1(c) and Supplemental Material [35]). The splitting of the (001) reflection at 0.82 Å$^{-1}$ indicates an incommensurate Dy ordering [Fig. 4(e)] with a modulation vector 0.0173(5) $c^*$. In the refinements, only transversal modulations of the $A_y$ component influence the splitting of the peak, and the amplitude of the transverse $M_y$ component is thus modulated by a sine-wave function with a periodicity of ~58 unit cells, corresponding to ~320 Å. Details for the refinement of the modulated structure can be found in the supplemental material [35]. Such modulation has also been reported for other orthoferrites [45, 46]. In orthomanganates, a splitting of a commensurate peak has also been reported in highly strained films, where a spin spiral coexists with an *E*-type ordering [47]. It should be noted also that the atomic positions for Dy are identical to the structural positions at higher temperature and do not show a shift as would be expected when an internal magnetic field gives rise to ferroelectricity.

### C. Electronic Structure

To gain further insights as to the mechanisms that may be responsible for the presence of a net magnetic component in the powder samples, we probed the electronic structure of the systems by means of x-ray absorption spectroscopy (XAS) at the Fe $L_{3,2}$, O K and Dy $M_{5,4}$ edge at room temperature for the powder and single crystal in total electron and fluorescence yield. For qualitative analysis, spectra taken with left and right circular polarized x-rays were averaged to yield isotropic spectra. Measuring both TEY and TFY gives us insight into surface related properties (TEY) whereas the fluorescence signal probes bulk properties [32, 33].

Comparing the spectra qualitatively for $M_{5,4}$ Dy of the powder and single crystal we find very little spectroscopic differences: peak positions, intensities and line shapes for TFY spectra are almost identical. The TEY intensity for the powder, however, is larger as compared to the crystal and probably related to differences in self-absorption between bulk and powder (not shown). These results are expected, since the $M_{5,4}$ edge probes transitions from $3d$ core-level into $4f$ final states, which do not play a leading role in the chemical bonding in the lanthanines. The situation for Fe is different (Fig. 5). Here we compare the $L_{3,2}$ TEY Fe spectrum of DyFeO$_3$ TEY and TFY powder and single crystal spectra with that of Fe$_2$O$_3$ as a reference since the Fe edge is expected to be more susceptible to changes in the oxygen content than the Dy edge. The peak position at 707.4 eV is identical for all spectra with very similar line shapes, indicating that the transition to the nonbonding Fe 3d $t_{2g}$ orbitals for the DyFeO$_3$ surface and bulk is the same and also the valence is +3. However, the position of the second peak at 708.9 eV, corresponding to the transitions to antibonding Fe 3d e*$_g$ states shows slight deviations. For the powder and single crystal, the TEY peak position is shifted by 0.1 eV with respect to Fe$_2$O$_3$ whereas the maximum for the single crystal in TFY is identical to the TEY powder peak position. The TFY powder signal is shifted in total by 0.35 eV to a larger energy. Since the TEY and TFY are almost identical for the single crystal and very similar for the powder, it is reasonable to assume that properties like oxidation state, oxygen content and structure for single crystal and powder are identical and homogenous. Fitting the line shape at 709 and 709.25 eV, we find that they are sharper as compared to Fe$_2$O$_3$ which indicates that the FeO$_6$ octahedra are less distorted and the energy for the crystal field splitting is larger. For the powder, there are other not fully identified contributions at higher energies. The shift in the peak position at 708.9 eV further indicates a small change of the Fe-O bond-length inside the DyFeO$_3$ grains, something which has not been picked up by the structural analysis. The small changes observed in the Fe-O bond-length in bulk or the nanocrystalline powder combined with the net magnetization of Dy could explain the observed FM in DyFeO$_3$ powders below $T_{SR}$. Indeed, the observed changes in the lattice constants with temperature are most likely responsible for the different magnetic symmetries found for powder and single crystal DyFeO$_3$ and for the magnetic behavior below $T_{SR}$, and hence provide an additional explanation as to why a net moment persists below $T_{SR}$.

### D. Discussion of Results

The results presented above provide evidence for a complex evolution of the magnetic and electronic states of DyFeO$_3$ in powder and single crystal forms with temperature. In particular, we show that the single crystal and powder samples exhibit strong differences in the Fe spin-reorientation temperature and different directions for the Fe and Dy spin reorientations. Since both crystalline forms exhibit similar lattice parameters at room temperature but different coefficients of thermal expansion, one may conclude that at the micrometer scale, new degrees of freedom are available that permit different magnetic equilibrium states, in particular, a lower Fe-Dy interaction of the order of 10 μeV (a factor of 10 lower than that for the single crystal) and the presence of an incommensurate Dy ordering at low temperatures.

Unexplained so far is the persistent magnetic component for the DyFeO$_3$ powder below $T_{SR}$ and why $\boldsymbol{m}_{Fe}$ for the powder below $T_{SR}$ points along [001] instead of the expected [010] observed for the single crystal. The magnetic field dependent single crystal neutron diffraction and magnetization measurements clearly show that small external stimuli have a large effect on the magnetic properties of DyFeO$_3$. A noticeable difference between powder and single crystal is the very different value for $T_{SR}$, ~56 K for the powder and 41 K for the crystal. Also, the *c*-axis lattice parameter for the single crystal is longer (7.658 Å) than the expected 7.623 Å [5]. These observations indicate that differences in the nearest-neighbour interaction, here Dy-Fe, between powder and crystal are responsible for the large difference in $T_{SR}$. If the interaction between Dy and Fe is weakened (elongation of the lattice), a lowering of $T_{SR}$ can be expected. This assumption is supported by determining the Dy-Fe distance along [001] from the reconstructed unit cells. The Dy-Fe distance for the powder is 3.088 Å whereas for the single crystal the distance is larger, 3.096 Å. The relative change in the lattice parameter between room temperature and 10 K for the powder is surprisingly small along the *a* and *b* directions, 0.09% and 0.16 %, respectively. This contrasts with the single crystal, where nearly identical changes for all unit cell paramters are observed between 300 K and 10 K, varying between 0.22 % and 0.26 %. Such asymmetry in the lattice parameter thermal expansion between powder and single cystal and the impact on the orbital overlap may be responsible, on the one hand, for the different Fe spins which flip at $T_{SR}$, from the [100] direction in both cases above $T_{SR}$ to the [010] direction for the single crystal and along [001] for the powder, and on the other hand to the different Dy spin-ordering planes, namely, the *ab*-plane for the crystal and the *bc*-plane for the powder. In the latter case, however, the FM component arising from the canting is not fully compensated, as is usually the case when going from from a $\Gamma_1$ ($\mathbf{G}_x\mathbf{A}_y\mathbf{F}_z$) to a $\Gamma_4$ ($A_x\mathbf{G}_y\mathbf{C}_z$) phase. These results demonstrate the strong susceptibility of DyFeO$_3$ to small purturbations in the magnetic energy, and in particular, to the role of the Fe-Dy interaction in defining the equilibrium state of the system.

**Conclusions**

We have studied the magnetic and structural properties of the orthoferrite DyFeO$_3$ as microcrystalline powder and bulk single crystal using x-ray and neutron diffraction, TEM electron diffraction, SQUID magnetometry, and XAS of the Dy $M_{5,4}$ and Fe $L_{3,2}$ edges at room temperature. Whereas the magnetic properties of the single crystal are largely consistent with previous reports, including for $T_{SR}$, $T_{N,Dy}$, and the Fe and Dy spin directions and ordering plane, the properties of the powder are distinct. The direction of the moment $m_{Fe}$ extracted for the powder is along [100] above $T_{SR}$ and along [001] below $T_{SR}$, whereas for the single crystal it is [100] above and [010] below $T_{SR}$. The Dy ordering plane for the powder is the *bc*-plane, while for the single crystal it is the *ab*-plane. In addition, we observe a coexisting incommensurate magnetic contribution in the powder with a modulation of 0.0173(5) *c**, corresponding to a modulation periodicity of ~58 unit cells. The $\Gamma_4$ ($A_x\mathbf{G}_y\mathbf{C}_z$) state of the single crystal below $T_{SR}$ fully compensates the FM contribution, whereas the $\Gamma_2$ ($\mathbf{F}_x C_y \mathbf{G}_z$) state of the powder results in partial compensation giving rise to a FM component that can be measured, absent in the single crystal.

The magnetic field dependent single crystal neutron diffraction measurements with $H\|c$ and $H\|a$ set an experimental energetic limit of approx. 10 μeV for the Fe-Dy interaction, whereas at least 100 μeV are needed to significantly overcome the Dy-Dy interaction. A similar value for the exchange interaction has been determine for $HoFeO_3$ (Fe-Ho: 26 μeV) by inelastic neutron scattering [48]. The experimentally estimated small energy value would explain why $DyFeO_3$ is very susceptible to external disturbances such as small magnetic fields. Our results also explain the observed differences in $T_{SR}$, since a smaller lattice constant strengthens the Fe-Dy interaction and increases $T_{SR}$ whereas an elongated lattice constant will result in opposite trends. This observation would also explain the partly reported differences in magnetic properties typically encountered for $DyFeO_3$ and likely for the orthoferrites more generally.

**Acknowledgments**

This work was supported by the Swiss National Science Foundation (Project No. 200020_169393) and the Paul Scherrer Institute. Part of this work was performed at the Swiss

Spallation Neutron Source, SINQ, the Surface/Interface Microscopy (SIM) and Materials Science (MS) beamline of the Swiss Light Source (SLS), Paul Scherrer Institut (PSI), Villigen, Switzerland. We also would like to thank Nicola Casati (PSI) for his help with the synchrotron powder measurements, Anja Weber (PSI) with the SEM measurements, and Simon Fluri (PSI) for his help with aligning and cutting the single crystal. We also would like to thank Frank Lichtenberg and Kaszimirz Conder for discussions on properties of single crystals and powders.

**Table I**

**Table caption:** Magnetic and structural data for DyFeO$_3$ single crystal and powder samples as determined by SQUID magnetometry and neutron diffraction measurements. The structural data taken at 10 K and 300 K for the single crystal are only to determine the relative change in lattice parameter. Numbers in parentesis are the standard deviation associated to the last digit.

|  | Powder (Neutrons) | Powder (XRD) | Single Crystal (Neutrons) | Single Crystal (XRD) | TEM (single crystal) |
|---|---|---|---|---|---|
| Lattice constants 300 K | $a$ = 5.309(8) Å $b$ = 5.602(2) Å $c$ = 7.632(5) Å | $a$ = 5.304(6) Å $b$ = 5.598(2) Å $c$ = 7.624(1) Å | $a$ = 5.304(2) Å $b$ = 5.597(1) Å $c$ = 7.604(5) Å | $a$ = 5.319(1) Å $b$ = 5.597(5) Å $c$ = 7.658(6) Å | $a$ = 5.381(9) Å $b$ = 5.561(5) Å $c$ = 7.636(9) Å |
| 10 K | $a$ = 5.304(4) Å $b$ = 5.593(2) Å $c$ = 7.616(9) Å |  | $a$ = 5.290(3) Å $b$ = 5.584(1) Å $c$ = 7.587(2) Å |  |  |
| Relative changes in lattice constants | Δ$a$=0.09 % Δ$b$=0.16 % Δ$c$=0.21 % |  | Δ$a$=0.26 % Δ$b$=0.23 % Δ$c$=0.22 % |  |  |
| $m_{Fe}$ below $T_{N,Fe}$ wFM | [100] [001] |  | [100] [001] |  |  |
| $T_{SR,Fe}$ $m_{Fe}$ below $T_{SR,Fe}$ | ~ 54 K [001] some FM |  | ~ 42 K [010] no wFM |  |  |
| Dy ordering $T_{N,Dy}$ ~ 4 K | $bc$-plane |  | $ab$-plane |  |  |

**Table II**

**Table caption:** Irreducible representations and ordering schemes for Fe (*4b*) and Dy (*4c*) in DyFeO$_3$ with space group Pbnm with $k$ = (0, 0, 0).

|            | Fe           | Dy           |
|------------|--------------|--------------|
| $\Gamma_1$ | $A_xG_yC_z$  | - - $C_z$    |
| $\Gamma_2$ | $F_xC_yG_z$  | $F_xC_y$ -   |
| $\Gamma_3$ | $C_xF_yA_z$  | $C_xF_y$ -   |
| $\Gamma_4$ | $G_xA_yF_z$  | - - $F_z$    |
| $\Gamma_5$ |              | $G_xA_y$ -   |
| $\Gamma_6$ |              | - - $A_z$    |
| $\Gamma_7$ |              | - - $G_z$    |
| $\Gamma_8$ |              | $A_xG_y$ -   |

**Figure captions:**

**Figure 1**.

a) Reconstructed DyFeO$_3$ unit cell from the powder diffraction data. b) Magnetic structure for DyFeO$_3$ powder between 450 and 100 K as derived from powder neutron diffraction measurements. Displayed is the Fe-lattice. c) Magnetic structure of the Fe-lattice below the spin reorientation reorientation for the powder. d) Magnetic structure of the ordered Dy-lattice at 2 K for the powder.

**Figure 2**.

a) Field cooled (FC) and zero field cooled (ZFC) magnetic moment measurements, $M(T)$, of DyFeO$_3$ powder taken with a magnetic field $H = 0.1$ T. b) ZFC measurements of the DyFeO$_3$ single crystal along [100], [010] and [001] with a magnetic field of $H = 0.01$ T. c) ZFC measurements of the DyFeO$_3$ single crystal along [001] with magnetic fields of $H = 0.01$, 0.02, and 0.1 T. d) FCC and ZFC measurements of the DyFeO$_3$ single crystal along [100] at different applied magnetic fields.

**Figure 3:**

$M(H)$ loops for DyFeO$_3$ powder taken at different temperatures (a) above and (b) below $T_{SR}$ and for the DyFeO$_3$ single crystal with the applied magnetic field along (c) [001] and (d) along [010].

**Figure 4:**

a) ZF temperature scan of the maximum peak intensity of the (011) and (031) Bragg peaks of the single crystal DyFeO$_3$ between 1.6 K and 70 K. Dashed line indicates $T_{SR}$. The inset shows the maximum peak intensity of the (011) Bragg peak between 35 K and 110 K. Temperature dependence of the maximum peak intensity the (031) / (03-1) single crystal Bragg peak for different magnetic field amplitudes applied alond (b) [100], (c) [010], and (d) [001]. e) Neutron powder diffraction patterns of DyFeO$_3$ powder collected at 2, 10, 100, 200, 300, and 450 K. The data shown are not corrected for absorption. f) Difference curve between 2 K and 10 K (top), and 10 K and 100 K (bottom) for powder neutron diffraction data of DyFeO$_3$ collected at DMC with $\lambda = 2.4575$ Å, giving the change in the magnetic contribution across $T_{N,Dy}$ and $T_{SR}$, respectively. The shaded regions indicate the positions of reflections from the aluminum sample holder.

**Figure 5:**

Normalized XAS spectra of the Fe $L_{3,2}$ absorption edge for Fe$_2$O$_3$ in total electron yield, and DyFeO$_3$ powder and single crystal in a) total electron and b) total fluorescence yield at room temperature.

**Figures:**

**Figure 1**

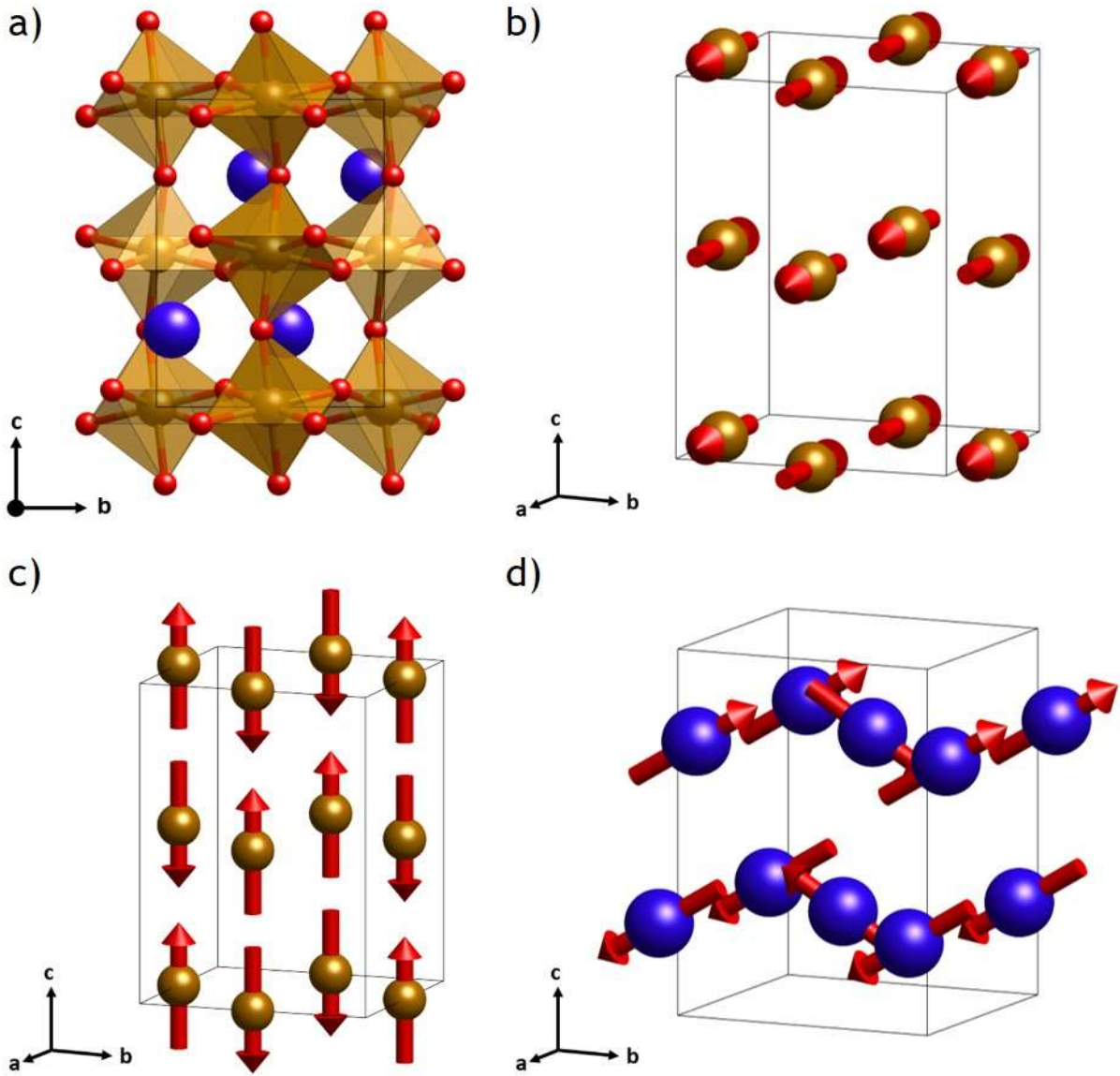

**Figure 2**

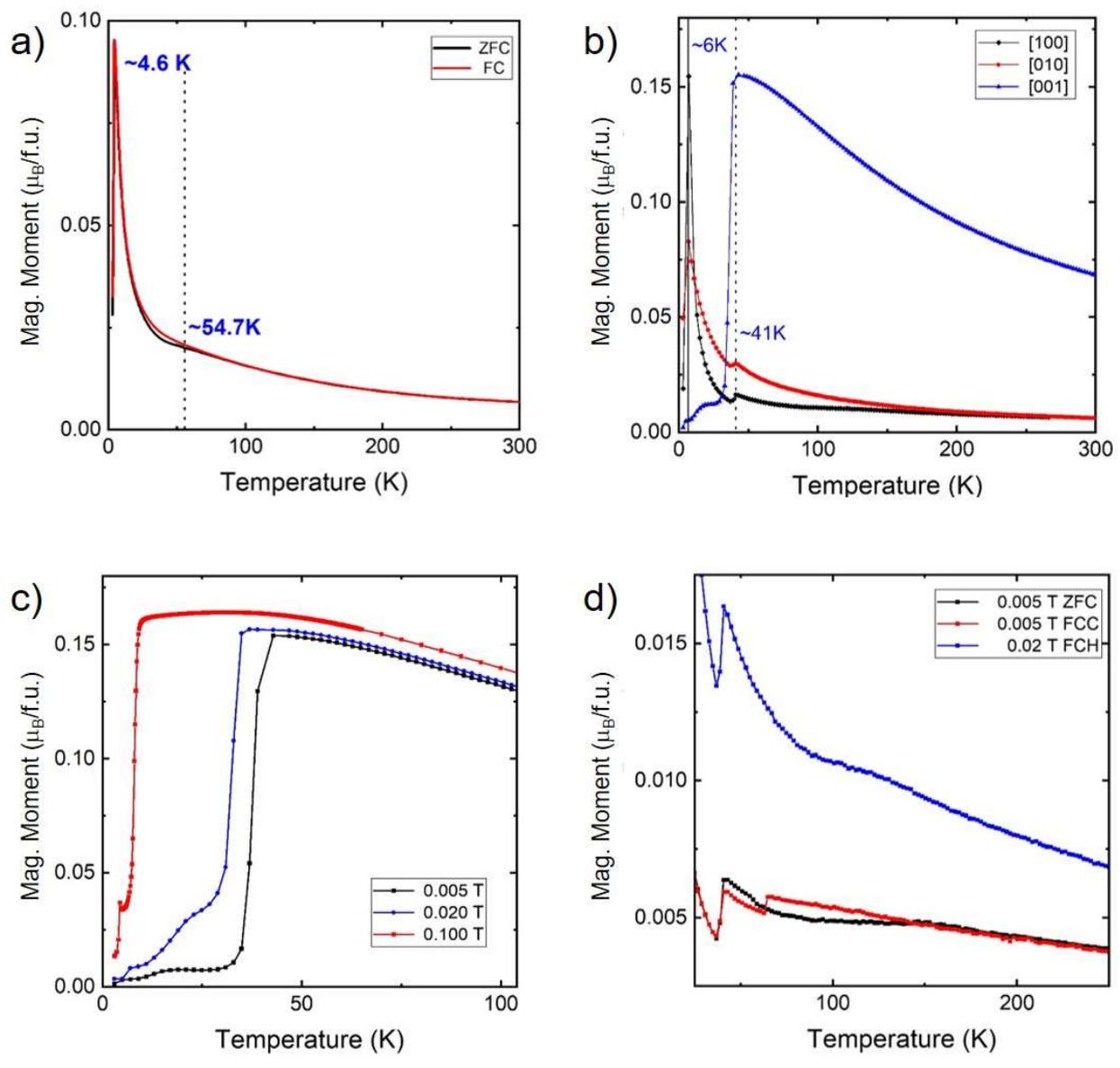

**Figure 3**

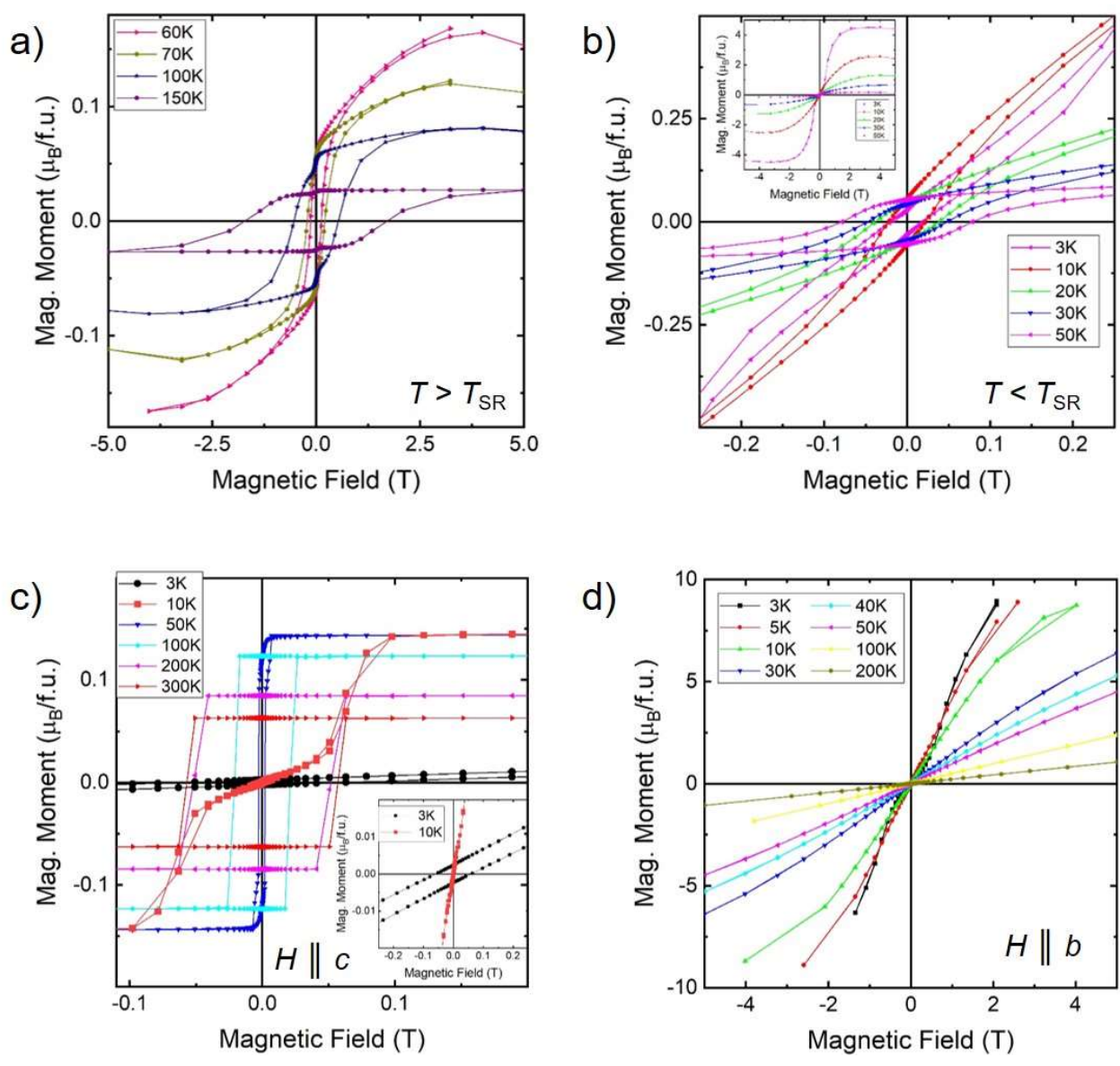

**Figure 4**

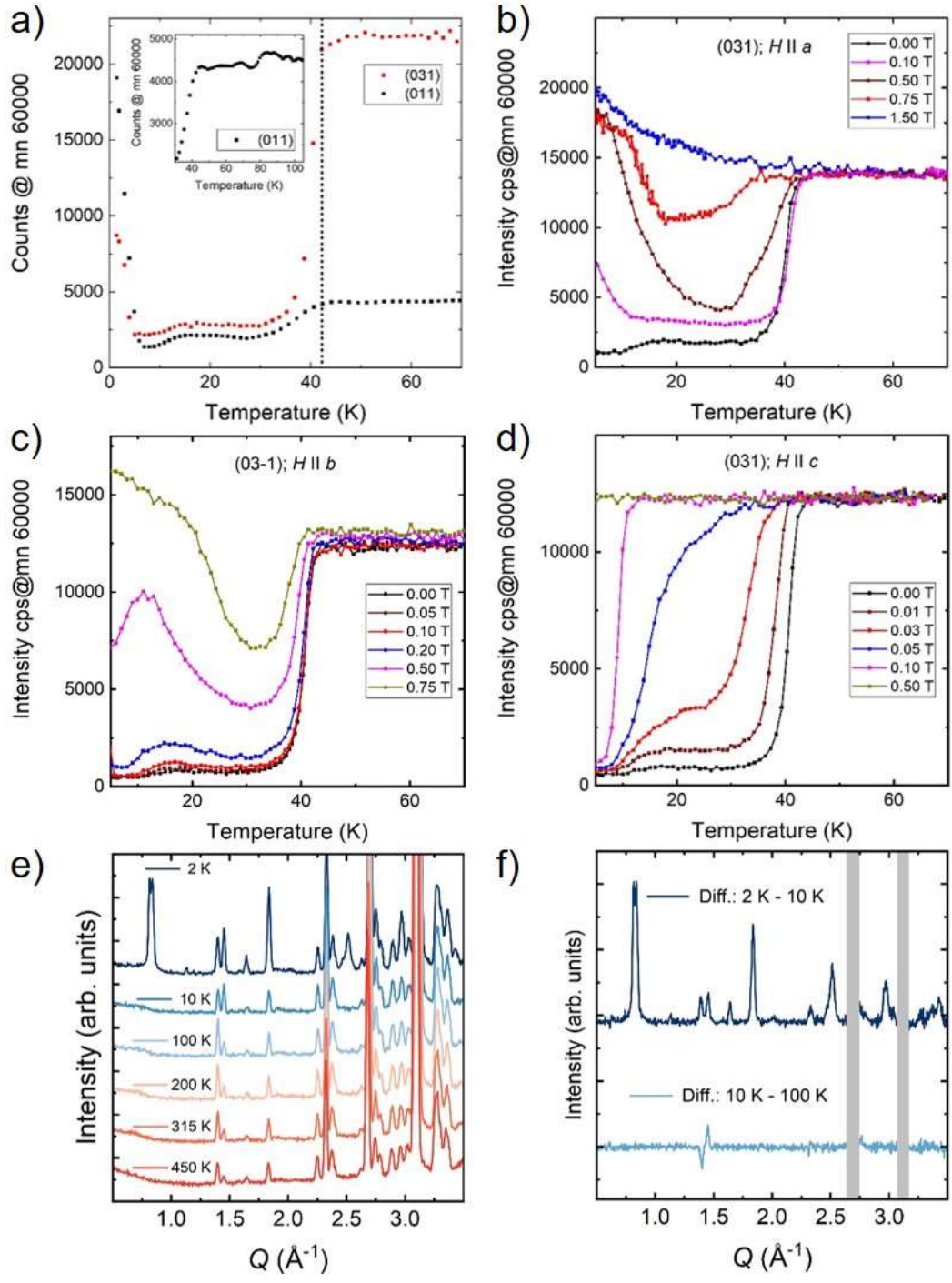

**Figure 5**

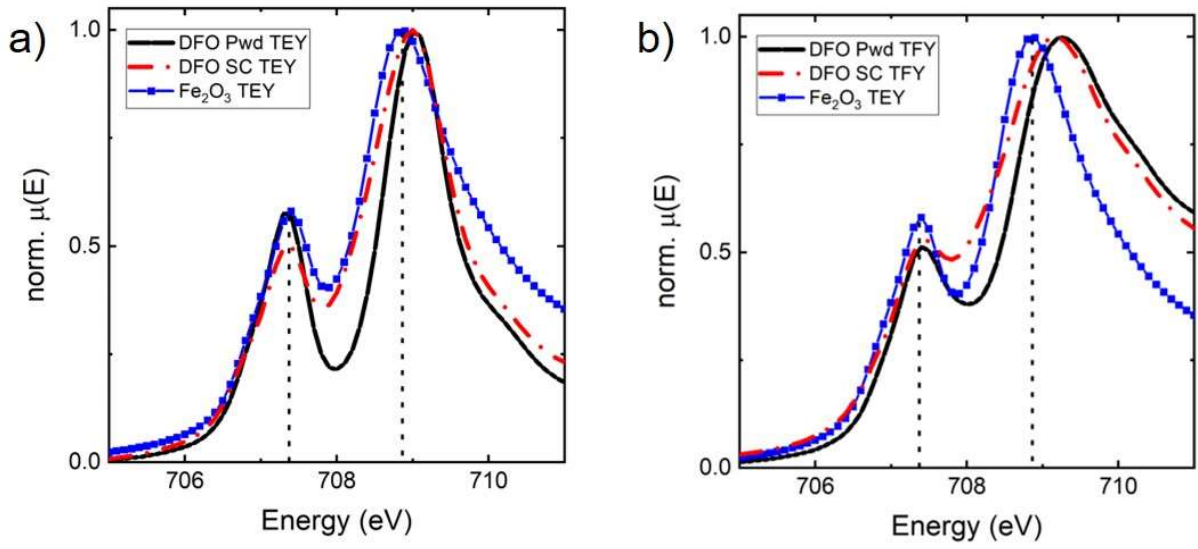

**Supplemental material for:**

# The role of Dy for magnetic properties of orthorhombic DyFeO$_3$

*Banani Biswas[1], Veronica F. Michel[1,5], Øystein S. Fjellvåg[1,2], Gesara Bimashofer[1], Max Döbeli[3], Michal Jambor[4], Lukas Keller[1], Elisabeth Müller[1], Victor Ukleev[1], Ekaterina V. Pomjakushina[1], Deepak Singh[1], Uwe Stuhr[1], Carlos A.F. Vaz[1], Thomas Lippert [1,5], and Christof W. Schneider,[1,]\**

[1] Paul Scherrer Institute, 5232 Villigen PSI, Switzerland
[2] Department for Hydrogen Technology, Institute for Energy Technology, NO-2027 Kjeller, Norway
[3] Ion Beam Physics, ETH Zurich, 8093 Zurich, Switzerland
[4] Institute of Physics of Materials, Czech Academy of Sciences, Czech Republic
[5] Department of Chemistry and Applied Biosciences, Laboratory of Inorganic Chemistry, ETH Zurich, Switzerland

**Refinement of powder neutron diffraction data at DMC:**

Powder neutron diffraction data for DyFeO$_3$ colleced at DMC was analyzed using the Jana2020 software [1]. The structural model from ref [2] (Table 1) was used for the nuclear model and lattice parameters, thermal displacement parameters, peak profile parameters (only Gaussian proved the best fit), and a seven term Legendre polynomial was used for background fitting. The zero shift was refined for one data set and the obtained value was used for all the datasets. The absorption from the sample was significant due to the absorption from Dy and the sample volume. The sample was not enriched with a less absorbing isotope. Due to the unknown density of the packed powder in the sample container, we could not perform a correction that would yield accurately corrected data. Thus the refined magnetic moments are not to be trused and the refined thermal displacement parameters are negative.

Table 1: Structural parameters in space group Pbnm used for analysis of neutron powder diffraction data collected at the DMC instrument at SINQ. Values obtained from ref [2].

| Site | Wyckoff position | $x$ | $y$ | $z$ | occ |
|---|---|---|---|---|---|
| Dy1 | 4c | -0.01725 | 0.06665 | 0.25000 | 1 |
| Fe1 | 4b | 0.50000 | 0.00000 | 0.00000 | 1 |
| O1 | 4c | 0.10810 | 0.46260 | 0.25000 | 1 |
| O2 | 8d | 0.30750 | -0.30330 | -0.05570 | 1 |

Magnetic refinements were carried out in Shubnikov group corresponding to the irreps of the magnetic ordering schemes. For ordering of Fe we carreied out the refinements for $\Gamma_4$ in Pb'n'm (M$_x$ dominated and M$_y$ and M$_z$ was set to zero), and $\Gamma_2$ in Pbn'm' (M$_z$ dominated and M$_x$ and M$_y$ was set to zero).

For the data 2K, we used $\Gamma_2$ for Fe and $\Gamma_5 + \Gamma_7$ for Dy. The refinements were carried out in the Shubnikov superspace group Pn'(a0g)0 to accommodate all irreps and the incommensurate ordering of the Dy moments. The reduction in symmetry generates more positions and coordinates for the structure are given in Table 2. At 2K, the moment of Fe was restricted to adopt the same values and orientation as at 10 K, assuming that the difference between 2 and 10 K purely originate from Dy ordering. Magnetic moments of Dy, the modulation vector and a sin component of the transversal modulations were refined. The Dy moment were restricted as: My[Dy2] = - My[Dy1], Mz[Dy2] = - Mz[Dy1], and Mysin[Dy2] = - Mysin[Dy1]. Refined magnetic moment for all temperatures are given in Table 3.

Table 2: Atomic positions of DyFeO$_3$ in Shubnikov superspace group Pn'(a0g)0.

| Site | x | y | z | occ |
|---|---|---|---|---|
| Dy1 | 0.01725 | 0.06665 | 0.25000 | 1 |
| Dy2 | -0.01725 | -0.06665 | -0.25000 | 1 |
| Fe1 | -0.50000 | 0.00000 | 0.00000 | 1 |
| Fe2 | -0.50000 | 0.00000 | 0.50000 | 1 |
| O1_1 | -0.10810 | 0.46260 | 0.25000 | 1 |
| O1_2 | 0.10810 | -0.46260 | -0.25000 | 1 |
| O2_1 | -0.30750 | -0.30330 | -0.05570 | 1 |
| O2_2 | -0.30750 | -0.30330 | 0.55570 | 1 |
| O2_3 | 0.30750 | 0.30330 | 0.05570 | 1 |
| O2_4 | 0.30750 | 0.30330 | 0.44430 | 1 |

Table 3: Refined magnetic moments. * not refined, locked to value at 10 K. For Dy at 2 K, Mysin was refined to 9.7(2) and the modulation vector to 0.0173(5) $c^*$. <u>NOTE</u>: DyFeO$_3$ absorbs neutron significantly and the refined values below cannot be trusted, as the data is not corrected for absorption.

| Temperature (K) | Fe | | | Dy | | |
|---|---|---|---|---|---|---|
| | Mx | My | Mz | Mx | My | Mz |
| 450 | 0 | 0 | 3.03(9) | 0 | 0 | 0 |
| 315 | 0 | 0 | 3.53(7) | 0 | 0 | 0 |
| 200 | 0 | 0 | 3.79(7) | 0 | 0 | 0 |
| 100 | 0 | 0 | 3.89(7) | 0 | 0 | 0 |
| 10 | 0 | 3.94(7) | 0 | 0 | 0 | 0 |
| 2 | 0 | 3.94* | 0 | 0 | 3.0(3) | 4.1(1) |

Table 4: R-factors from Rietveld refinements of powder neutron diffraction data of DyFeO$_3$ measured at DMC.

| Temperature (K) | GOF | Rp | Rwp |
|---|---|---|---|
| 450 | 1.86 | 5.05 | 6.94 |
| 315 | 1.69 | 5.16 | 6.48 |
| 200 | 1.73 | 5.46 | 6.82 |
| 100 | 1.73 | 5.44 | 6.89 |
| 10 | 1.69 | 5.51 | 6.77 |
| 2 | 2.38 | 7.44 | 9.35 |

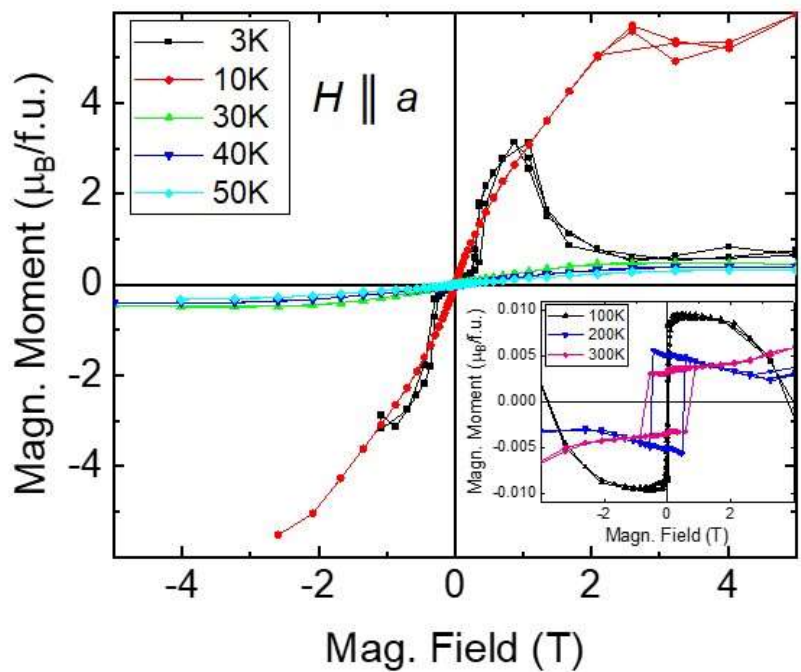

Figure S1: $M(H)$ loops for the DyFeO$_3$ single crystal taken at different temperatures with the applied magnetic field along the [100] direction.

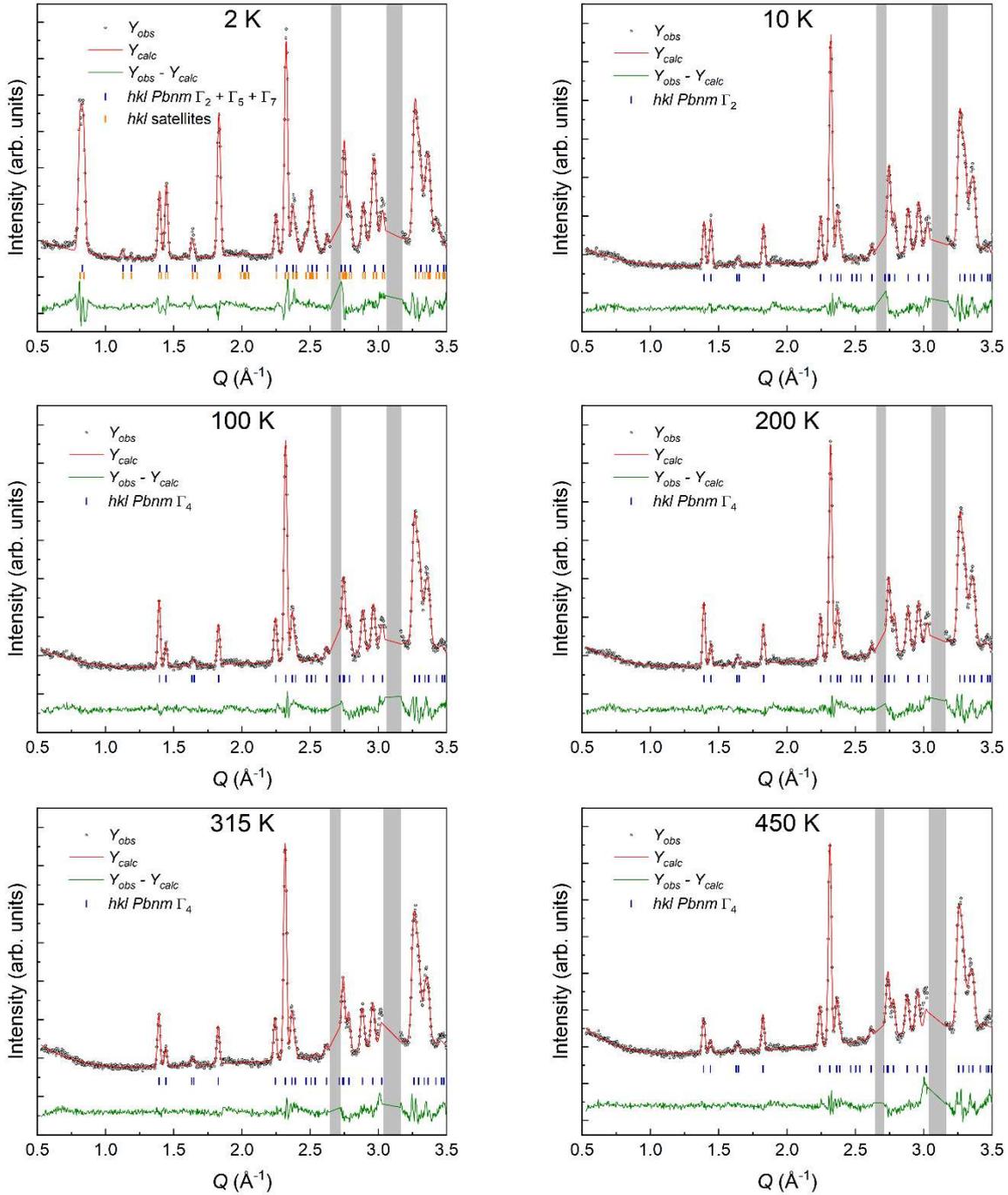

Figure S2: Measured, calculated and difference profiles for Rietveld refinement of the magnetic structure of DyFeO$_3$ from neutron diffraction data recorded on DMC between 2 and 450 K with $\lambda$ = 2.4575 Å. Blue tics indicate reflections from the magnetic structure. 2 K: Pbnm with $\Gamma2 + \Gamma5 + \Gamma7$ (refined in Pn'). 10 K: Pbnm with $\Gamma2$ (refined in Pbn'm'). 100-450 K: Pbnm with $\Gamma4$ (refined in Pb'n'm). At 2 K the orange tics indicate satellite reflections from the modulation. The shaded regions indicate the positions of reflections from the aluminum sample can and are excluded from the refinements.
36